\documentclass[superscriptaddress,preprintnumbers,nofootinbib,showpacs]{revtex4}

\usepackage{graphicx}
\usepackage{latexsym}
\usepackage[english]{babel}
\usepackage{amsmath}
\usepackage{amsthm}
\usepackage{epsfig}
\usepackage{color}
\usepackage{amssymb}

\begin{document}

\title{\Large The Scalar Field Kernel in Cosmological Spaces}

\preprint{ITP-UU-07/65, SPIN-07/50}

\preprint{HIP-2008-02/TH}


\author{Jurjen F. Koksma}
\email[]{J.F.Koksma@phys.uu.nl}
\affiliation{Institute for
Theoretical Physics (ITP) \& Spinoza Institute, Utrecht
University, Postbus 80195, 3508 TD Utrecht, The Netherlands}

\author{Tomislav Prokopec}
\email[]{T.Prokopec@phys.uu.nl}
\affiliation{Institute for
Theoretical Physics (ITP) \& Spinoza Institute, Utrecht
University, Postbus 80195, 3508 TD Utrecht, The Netherlands}

\author{Gerasimos I. Rigopoulos}
\email[]{gerasimos.rigopoulos@helsinki.fi}
\affiliation{Helsinki
Institute of Physics, University of Helsinki, P.O. Box 64,
FIN-00014, Finland}

\begin{abstract}
We construct the quantum mechanical evolution operator in the
Functional Schr\"{o}dinger picture \mbox{-- the kernel --} for a
scalar field in spatially homogeneous FLRW spacetimes when the
field is a) free and b) coupled to a spacetime dependent source
term. The essential element in the construction is the causal
propagator, linked to the commutator of two Heisenberg picture
scalar fields. We show that the kernels can be expressed solely in
terms of the causal propagator and derivatives of the causal
propagator. Furthermore, we show that our kernel reveals the
standard light cone structure in FLRW spacetimes. We finally apply
the result to Minkowski spacetime, to de Sitter spacetime and
calculate the forward time evolution of the vacuum in a general
FLRW spacetime.
\end{abstract}

\maketitle

\section{Introduction}
\label{Introduction}

The Functional Schr\"odinger picture is based on the projection of
quantum states and operators on the field amplitude basis. Guth
and Pi \cite{Guth:1985ya} have already used it in the study of
inflationary perturbations and related the width of the Gaussian
vacuum wave functional in de Sitter spacetime to Heisenberg
picture scalar fields, thus casting quantum field theory in terms
familiar from non-relativistic quantum mechanics.

A kernel is a quantum mechanical evolution operator in the
Functional Schr\"odinger picture and is the fundamental object of
any quantum mechanical theory.\footnote{Note that in some of the
literature this object is referred to as the quantum mechanical
propagator.} It represents a transition amplitude from an
arbitrary initial state at time $t'$ to an arbitrary final state
at time $t$. Moreover, it allows to calculate the forward time
evolution of any given initial field configuration. In this paper
we demonstrate that the kernel can be expressed solely in terms of
the causal propagator and derivatives of the causal propagator
thus elaborating further on the connection with the Heisenberg
(operator) picture for quantum field theories. Furthermore, the
appearance of the causal propagator makes the causality structure
of the theory evident.

The kernel allows for many physical applications in quantum field
theory. For example, if interactions between various scalar fields
or fermionic fields are linear in (one of the) matter fields, the
formalism developed here can straightforwardly be applied.
Furthermore, another natural application of the kernel can be
found in the study of decoherence \cite{Prokopec:2006fc,
Koksma:2007zz} accounting for the quantum-to-classical transition
in the early Universe.

After having reviewed some basic quantum mechanics in the current
section, we will calculate the kernel for a free scalar field in
section \ref{The Kernel in the Non-Interacting Case}, and a scalar
field coupled to a source term in section \ref{The Kernel in the
Interacting Case} in cosmological spacetimes (spatially
homogeneous backgrounds). In section \ref{Causality in Quantum
Mechanics} we elaborate on causality in quantum field theory
expressed in the Functional Schr\"odinger picture and end in
section \ref{Applications} with some examples, to wit, the simple
harmonic oscillator, the kernel in de Sitter spacetime and the
time evolution of the vacuum state in
Friedmann-Lema\^itre-Robertson-Walker or FLRW spacetimes.

\subsection{Essentials from Quantum Mechanics}
\label{Essentials from Quantum Mechanics}

Let us begin by recalling some basic identities from quantum
mechanics. A quantum state $|\Psi(t)\rangle$ at time $t$ can in
general be expressed in terms of the evolution operator
$\hat{U}(t,t')$ and the quantum state $|\Psi(t')\rangle$ at
initial time $t'$ as follows:
\begin{equation}\label{evolution}
|\Psi(t)\rangle = \hat{U}(t,t')|\Psi(t')\rangle \,.
\end{equation}
Just as the state $|\Psi(t)\rangle$, the evolution operator
$\hat{U}(t,t')$ obeys the Schr\"odinger equation:
\begin{equation}\label{Schrodinger}
i\hbar\frac{\partial}{\partial t}\hat{U}(t,t') = \hat H \,
 \hat{U}(t,t')
\,,
\end{equation}
where $\hat{H}$ is the Hamiltonian operator of the system under
consideration. The formal solution is a time ordered (${\rm T}$)
exponential:
\begin{equation}\label{evolution:solution}
\hat{U}(t,t')= \mathrm {T}\, \exp{\left(-\frac{i}{\hbar}
\int^{t}_{t'} dt'' \hat{H}(t'') \right)} \,.
\end{equation}
Note that $t > t'$ is implied and when $t < t'$ time ordering
should be replaced with anti-time ordering ($\overline{{\rm T}}$).
From this equation we can easily infer some important properties
the evolution operator satisfies:
\begin{subequations}
\label{evolutionproperties}
\begin{eqnarray}
\hat U^{\dag}(t,t') &=& \hat U(t',t)
\label{evolutionproperties1} \\
\hat{U}(t',t') &=& 1
\label{evolutionproperties2} \\
\hat{U}(t,t'')\hat{U}(t'',t') &=&
\hat{U}(t,t')\label{evolutionproperties3} \,.
\end{eqnarray}
\end{subequations}

\subsection{Expectation Values and Commutators}
\label{Equivalence Relation for Expectation Values}

Let us now state a relation between quantum mechanical expectation
values and commutators (see for example \cite{Weinberg:2005vy})
which will play an important r\^ole in the interpretation of the
kernel. We consider the expectation value of some general operator
$\hat{Q}$ in the Schr\"odinger picture:
\begin{equation}\label{expectationvalueQ1}
\langle \hat{Q}(t)\rangle \equiv
\langle\Psi(t)|\hat{Q}|\Psi(t)\rangle \,.
\end{equation}
Now, by using (\ref{evolution:solution}) for the evolution
operator:
\begin{equation}\label{expectationvalueQ2}
\langle \hat{Q}(t)\rangle=\left\langle\Psi(t')\left|\left\{
\overline{\mathrm {T}}\, \exp{\left(\frac{i}{\hbar} \int^{t}_{t'}
dt'' \hat{H}(t'') \right)}\right\} \hat{Q} \left\{\mathrm {T}\,
\exp{\left(-\frac{i}{\hbar} \int^{t}_{t'} dt'' \hat{H}(t'')
\right)}\right\}\right| \Psi(t')\right\rangle \,,
\end{equation}
one can prove by induction that at each order in $\hat{H}$ one has:
\begin{equation}\label{expectationvalueQ3}
\langle \hat{Q}(t)\rangle=
\sum_{n=0}^{\infty}\left(\frac{i}{\hbar}\right)^{n}
\int_{t'}^{t}dt_{n}\int_{t'}^{t_{n}}dt_{n-1} \cdots
\int_{t'}^{t_{2}}dt_{1}\left\langle\Psi(t')\left|
\left[\hat{H}(t_{1}) ,\left[\hat{H}(t_{2}), \cdots
\left[\hat{H}(t_{n}),\hat{Q}\right]
\cdots\right]\right]\right|\Psi(t')\right\rangle.
\end{equation}
We can conclude that expectation values, naturally defined in
terms of evolution operators as in (\ref{expectationvalueQ2}), can
in a fully equivalent manner be calculated from the expectation
value of nested commutators, as in (\ref{expectationvalueQ3}).
Information about observables is hence equally well stored in a
series of commutators as in evolution operators.

In quantum mechanics causality corresponds to the statement that
all commutators of observables, including non-commuting
observables, vanish outside past and future light cones.
Alternatively, measurements performed at points in spacetime with
a spacelike separation can be carried out simultaneously. This
equivalence relation already suggests that quantum mechanics is a
causal theory in full generality since, up to higher order
irreducible $n$-point functions, expectation values can be
expressed in terms of the commutator, a manifestly causal
quantity. These are vital observations, and we will return to them
shortly.

\subsection{The Causal Propagator}
\label{The Causal Propagator}

Given some quantum field operator $\hat{\phi}(x)$ in the
Heisenberg picture one can construct various vacuum expectation
values by means of the Schwinger-Keldysh formalism (see e.g.
\cite{Prokopec:2003tm}). We will be interested in a particular
linear combination of the two Wightman functions ($G_{+-}$ and
$G_{-+}$) to construct what we will henceforth refer to as the
causal propagator:
\begin{equation} \label{commutatorpropagator}
G_{c}(x,x') \equiv G_{-+}(x,x')-G_{+-}(x,x') = \left\langle
\Omega\left|\left[\hat{\phi}(x),\hat{\phi}(x')\right]\right|
\Omega\right\rangle \equiv \left\langle\Omega\left|\hat{\phi}(x)
\hat{\phi}(x') - \hat{\phi}(x') \hat{\phi}(x)
\right|\Omega\right\rangle \,.
\end{equation}
In particular, if we assume a spatially homogeneous background and
a quadratic Hamiltonian, expanding in terms of creation and
annihilation operators:
\begin{equation} \label{expansioncontphi}
\hat{\phi}(x) = \int \frac{d^{3}\mathbf{k}}{(2\pi)^{3}}
\hat{a}_{\mathbf{k}}\,\phi_{k}(t)e^{i\mathbf{k\cdot x}} +
\hat{a}_{\mathbf{k}}^{\dag}\,\phi_{k}^{\ast}(t)e^{-i\mathbf{k\cdot
x}}\,,
\end{equation}
yields:
\begin{equation} \label{commutatorpropagatorFourierspace}
G_{c}(x,x') = \hbar \int\frac{d^{3}\mathbf{k}}{(2\pi)^{3}}
G_{c}(k,t,t')e^{i\mathbf{k}\cdot(\mathbf{x}-\mathbf{x}')} \,,
\end{equation}
where:
\begin{equation} \label{commutatorpropagatorF}
G_{c}(k,t,t') =  \phi_{k}(t)\phi_{k}^{\ast}(t')-
\phi_{k}^{\ast}(t)\phi_{k}(t')  \,.
\end{equation}
Here, the $\hbar$ originates from imposing the standard
commutation relations between creation and annihilation operators.
Furthermore, note the field modes are homogeneous, i.e.:
$\phi_{k}(t)$ depends on $k = \|\mathbf{k}\|$.

This propagator is causal in the quantum mechanical sense because
it originates from the commutator~\cite{Peskin:1995ev},
unlike for example the Feynman
or (anti-)time ordered propagators.

\subsection{The Functional Schr\"odinger Picture}
\label{The Functional Schrodinger Picture}

In the Functional Schr\"odinger picture
\cite{Jackiw:1988sf,Jackiw:1987aq,Guth:1985ya, Guven:1987bx} a
quantum mechanical state $|\Psi(t)\rangle$ is realised by a wave
functional $\Psi(\phi,t)$ which is a functional of the $c$-number
functions $\phi=\{\phi(\mathbf{x}), \forall\,\mathbf{x} \in
\mathrm{R}^{3}\}$ defined by the projection on the field amplitude
basis $\Psi(\phi,t)=\langle\phi|\Psi(t)\rangle$, where
$|\phi\rangle=\prod_{\mathbf{x}}|\phi(\mathbf{x})\rangle$. The
action of a quantum field operator $\hat{\phi}(\mathbf{x})$ and
its associated canonical momentum $\hat{\pi}(\mathbf{x})$ are
given by:
\begin{eqnarray}
&& \langle\phi|\hat{\phi}(\mathbf{x}) |\Psi(t)\rangle =
\phi(\mathbf{x}) \Psi(\phi,t) \label{phiaction}  \\
&& \langle\phi|\hat{\pi}(\mathbf{x}) |\Psi(t)\rangle =
\frac{\hbar}{i} \frac{\delta}{\delta \phi(\mathbf{x})}
\Psi(\phi,t)\label{piaction}\,.
\end{eqnarray}

\section{The Kernel for the free scalar field}
\label{The Kernel in the Non-Interacting Case}

Let us now examine the quantum mechanical
evolution operator (\ref{evolution:solution}) in the Functional
Schr\"odinger picture:
\begin{equation}\label{projection-functionalintegral}
K(\phi,t;\phi',t') \equiv \langle \phi|\hat{U}(t,t')|\phi'\rangle
= \int_{\phi''(t')=\phi'}^{\phi''(t)=\phi}
\mathcal{D}\phi''\exp\left [ \frac{i}{\hbar}\,S[\phi'']\right ]
\,.
\end{equation}
In the above equation, $K(\phi,t;\phi',t')$ is the so-called
kernel, a transition amplitude from some initial state $\phi'$ at
$t'$ to the state $\phi$ at $t$. The kernel can be expressed in
terms of a path integral of the action where we integrate over all
intermediate field configurations.

It is interesting to note that from properties
(\ref{evolutionproperties}) we deduce the following symmetry
requirements for $K$:
\begin{subequations}
\label{symreqK}
\begin{eqnarray}
K^{\ast}(\phi,t;\phi',t') &=& K(\phi',t';\phi,t) \label{symreqK1}
\\
K(\phi,t;\phi',t) &=& \delta(\phi-\phi') \label{symreqK2} \\
K(\phi,t;\phi',t') &=& \int \mathcal{D}\phi''
K(\phi,t;\phi'',t'')K(\phi'',t'';\phi',t') \label{symreqK3} \,,
\end{eqnarray}
\end{subequations}
where the functional delta function has to be understood as:
$\delta(\phi-\phi') = \prod_{\mathbf{x}}
\delta(\phi(\mathbf{x})-\phi'(\mathbf{x}))$.

The action for a real scalar field $\phi(x)$ for a finite time
interval ranging between $t'$ and $t$ generally valid for real
quantum fields in curved spacetimes is given by:
\begin{equation} \label{action}
S[\phi] = \int d^{4}x \, \sqrt{-g} \left( -\frac{1}{2}\,
\partial_{\alpha} \phi(x) \, \partial_{\beta} \phi(x) \, g^{\alpha\beta}
-\frac{1}{2}\,(m^{2}+\xi R)\phi^{2}(x)\right )\,,
\end{equation}
where $R$ denotes the Ricci curvature scalar and $g = {\rm det}
[g_{\mu\nu}]$. Let us specialise to FLRW spacetimes in which the
metric is given by $g_{\alpha\beta}= \mathrm{diag}
\left(-1,a^{2}(t),a^{2}(t),a^{2}(t)\right)$ where $a(t)$ is the
scale factor of the Universe. Classically, this action leads to
the standard equation of motion:
\begin{equation} \label{motionphi}
\Box \phi_{\mathrm{cl}}(x)-(m^{2}+\xi R)\phi_{\mathrm{cl}}(x)=0
\,,
\end{equation}
where $\Box = (-g)^{-1/2}\partial_\mu (-g)^{1/2}g^{\mu\nu}
\partial_\nu$ is the scalar d'Alambertian. We write:
\begin{equation}\label{qfielcfield}
\phi(x)=\phi_{\mathrm{cl}}(x)+\delta\phi(x)\,,
\end{equation}
and insert this into equation (\ref{action}). The boundary
conditions on $\phi(x)$ are carried by the classical field only,
i.e.: $\phi_{\mathrm{cl}}(\mathbf{x},t')=\phi'(\mathbf{x})$ and
$\phi_{\mathrm{cl}}(\mathbf{x},t)=\phi(\mathbf{x})$, and
straightforwardly result into the requirement that
$\delta\phi(\mathbf{x},t')=0=\delta\phi(\mathbf{x},t)$. Assuming
that the classical field (\ref{motionphi}) vanishes at spatial
infinity, two straightforward partial integrations yield:
\begin{equation}\label{actionrewrite1}
S[\phi_{\mathrm{cl}}+\delta\phi] = S[\delta\phi] +
S[\phi_{\mathrm{cl}}] = S[\delta\phi] + \frac{1}{2} \int
d^{3}\mathbf{x} \,\phi_{\mathrm{cl}}(x)\, \pi_{\mathrm{cl}}(x)
\Big|_{t'}^{t} \,,
\end{equation}
where we have recognised the canonical momentum associated to
$\phi_{\mathrm{cl}}(x)$ given by $\pi_{\mathrm{cl}}(x)=
a^{3}(t)\,\dot{\phi}_{\mathrm{cl}}(x)$. Equation
(\ref{projection-functionalintegral}) boils down to the following
result:
\begin{equation} \label{kernel1}
K(\phi,t;\phi',t') = \exp\left[\frac{i}{2\hbar} \int
d^{3}\mathbf{x}  \phi_{\mathrm{cl}}(x)\, \pi_{\mathrm{cl}}(x)
\Big|_{t'}^{t}\right] \int_{\delta\phi(t')=0}^{\delta\phi(t)=0}
\mathcal{D}\delta\phi\,\,\exp\left [
\frac{i}{\hbar}\,S[\delta\phi]\right ]\,.
\end{equation}
Note that the remaining path integral represents the transition
amplitude between zero field states and is thus independent of the
original boundary conditions on the field $\phi(x)$. Hence we can
absorb this factor into the overall normalisation of the kernel:
\begin{equation} \label{kernel2}
K(\phi,t;\phi',t') = \mathcal{M}(t,t') \exp\left[\frac{i}{2\hbar}
\int d^{3}\mathbf{x}  \phi_{\mathrm{cl}}(x)\, \pi_{\mathrm{cl}}(x)
\Big|_{t'}^{t}\right]\,.
\end{equation}
Next, we Fourier expand the classical field:
\begin{equation} \label{fieldfouriertransform}
\phi_{\mathrm{cl}}(x)=\int \frac{d^{3}\mathbf{k}}{(2\pi)^{3}}\,
\phi_{\mathbf{k}}(t)\,\mathrm{e}^{i\mathbf{k}\cdot\mathbf{x}}\,,
\end{equation}
which is well defined because we consider a spatially flat FLRW
background. Note the property
$\phi^{\ast}_{\mathbf{k}}(t)=\phi_{-\mathbf{k}}(t)$ and the
omission of the subscript ``cl" of the field modes for future
convenience. We must keep in mind though, that also the field
modes are classical in the sense that they obey the Fourier
transform of equation of motion (\ref{motionphi}):
\begin{equation} \label{motionphi2}
\left ( \frac{\partial^{2}}{\partial {t}^{2}}+3H\frac{\partial}
{\partial t}+\frac{k^{2}} {a^{2}(t)}+m^{2}+\xi R
\right)\phi_{\mathbf{k}}(t)=0\,,
\end{equation}
where $H(t)=\dot{a}(t)/a(t)$ is the Hubble parameter. We thus
have:
\begin{equation}\label{actionrewrite2}
S[\phi_{\mathrm{cl}}] = \left.\frac{1}{4} \, \int
\frac{d^{3}\mathbf{k}}{(2\pi)^{3}}\, \left[a^{3}(\tilde{t}\,)
\partial_{\tilde{t}}\left | \phi_{\mathbf{k}} (\tilde{t}\,)\right
|^{2}\right]\right|_{\tilde{t}=t'}^{\tilde{t}=t} \,.
\end{equation}
Note that running time $\tilde{t}$ is defined on the interval
$t'\leq \tilde{t} \leq t$. Let us denote the two fundamental
solutions of (\ref{motionphi2}) by
$\chi_{\mathbf{k}}(\tilde{t}\,)$ and
$\chi_{\mathbf{k}}^{\ast}(\tilde{t}\,)$. In general,
$\phi_{\mathbf{k}}(\tilde{t}\,)$ is a linear superposition of the
two fundamental solutions, i.e.: $\phi_{\mathbf{k}}(\tilde{t}\,) =
\alpha_{\mathbf{k}} \,\chi_{\mathbf{k}}(\tilde{t}\,) +
\beta_{\mathbf{k}}\, \chi_{\mathbf{k}}^{\ast}(\tilde{t}\,)$. We
are now in the position to impose the boundary conditions at
initial and final times and solve for $\alpha_{\mathbf{k}}$ and
$\beta_{\mathbf{k}}$. We thus find:
\begin{equation}\label{phisolutionpropagator}
\phi_{\mathbf{k}}(\tilde{t}\,) = \phi_{\mathbf{k}}
\frac{G_{\mathbf{k}}(\tilde{t},t')}{G_{\mathbf{k}}(t,t')} +
\phi'_{\mathbf{k}}
\frac{G_{\mathbf{k}}(t,\tilde{t}\,)}{G_{\mathbf{k}}(t,t')}\,,
\end{equation}
where $G_{\mathbf{k}}(t,t')$ is related to the causal propagator
(\ref{commutatorpropagatorF}) as follows:
\begin{equation}\label{causalproplinkGprop}
G_{\mathbf{k}}(t,t') =
\phi_{\mathbf{k}}(t)\phi_{\mathbf{k}}^{\ast}(t')-
\phi_{\mathbf{k}}^{\ast}(t)\phi_{\mathbf{k}}(t') =
G_{c}(k,t,t')\mathrm{sgn}(\mathbf{k}\cdot\hat{\mathbf{n}})\,,
\end{equation}
where $\hat{\mathbf{n}}$ is a unit vector in k-space normal to an
arbitrary plane through the origin. The sign-function is
introduced in order to preserve the odd symmetry under
$\mathbf{k}\rightarrow -\mathbf{k}$. One can thus express the
solution to differential equation (\ref{motionphi2}) in terms of
its boundary conditions and causal propagators exclusively. We
believe that this is also true for general spacetimes and fields
of non-zero spin~\cite{ProkopecRigopoulos:2007}.

Note that the Wronskian yields:
\begin{equation}\label{Wronksiandefinition}
W[\phi_{\mathbf{k}}(t), \phi_{\mathbf{k}}^{\ast}(t)] \equiv
\phi_{\mathbf{k}}(t) \dot{\phi}^{\ast}_{\mathbf{k}}(t) -
\phi^{\ast}_{\mathbf{k}}(t) \dot{\phi}_{\mathbf{k}}(t) =
\left\{|\alpha_{\mathbf{k}}|^{2}- |\beta_{\mathbf{k}}|^{2}\right\}
W[\chi_{\mathbf{k}}(t), \chi_{\mathbf{k}}^{\ast}(t)]\,,
\end{equation}
where:
\begin{equation}\label{Wronskianphi03}
|\alpha_{\mathbf{k}}|^{2}-
|\beta_{\mathbf{k}}|^{2}=\frac{1}{G_{\mathbf{k}}(t,t')}\Big(
\phi_{\mathbf{k}} {\phi'}^{\ast}_{\mathbf{k}}
-\phi^{\ast}_{\mathbf{k}}\phi'_{\mathbf{k}}\Big) \,.
\end{equation}
Generally, this equation does not equal unity, a requirement
usually imposed for a consistent canonical quantisation in the
Heisenberg picture, because the boundaries are arbitrary. Indeed,
this is an important difference between the Schr\"odinger and
Heisenberg pictures.

Next, we substitute (\ref{phisolutionpropagator}) into our
modified action, equation (\ref{actionrewrite2}). The finite
volume (in position space) result for the kernel in Fourier space
reads:
\begin{equation}\label{PropKernel}
K(\phi,t;\phi',t') =
\prod_{\mathbf{k}}\,\mathcal{M}_{\mathbf{k}}\,  \exp\Bigg[
-\frac{1}{V}\Big\{B(k,t,t')\,
\phi_{\mathbf{k}}\phi^{\ast}_{\mathbf{k}} +
C(k,t,t')\,\phi'_{\mathbf{k}}{\phi'}^{\ast}_{\mathbf{k}}
 +\frac{1}{2}D(k,t,t')\,
\Big(\phi_{\mathbf{k}}{\phi'}^{\ast}_{\mathbf{k}} +
\phi^{\ast}_{\mathbf{k}}{\phi'}_{\mathbf{k}}\Big)\Big \} \Bigg]
\,,
\end{equation}
where $\mathcal{M}_{\mathbf{k}}$ is a normalisation constant and
where:
\begin{subequations}
\label{PropKernelconst}
\begin{eqnarray}
B(k,t,t') &=& -\frac{ia^{3}(t)}{2\hbar}\,
\frac{\partial_{t}G_{\mathbf{k}}(t,t')} {G_{\mathbf{k}}(t,t')}
\label{PropKernelconstB} \\ C(k,t,t') &=&
\frac{ia^{3}(t')}{2\hbar}\,
\frac{\partial_{t'}G_{\mathbf{k}}(t,t')} {G_{\mathbf{k}}(t,t')}
\label{PropKernelconstC} \\ D(k,t,t') &=& -\frac{i}{2\hbar
G_{\mathbf{k}}(t,t')}\Big\{ a^{3}(t) W[\chi_{\mathbf{k}}(t),
\chi_{\mathbf{k}}^{\ast}(t)] + a^{3}(t') W[\chi_{\mathbf{k}}(t'),
\chi_{\mathbf{k}}^{\ast}(t')] \Big\} \label{PropKernelconstD}\,.
\end{eqnarray}
\end{subequations}
Note that $B(\mathbf{k},t,t')= B(k,t,t')$ due to the homogeneity
of space. The volume factor appearing in (\ref{PropKernel}) is due
to the identification $\int\frac{d^{3}\mathbf{k}}{(2\pi)^{3}} =
\frac{1}{V}\sum_{\mathbf{k}}$, valid in the infinite volume limit,
which we use throughout this paper. In FLRW spacetimes, the
Wronskian equals:
\begin{equation}\label{wronskian}
W[\chi_{\mathbf{k}}(t), \chi_{\mathbf{k}}^{\ast}(t)]
=\frac{i}{a^{3}(t)} \mathrm{sgn}(\mathbf{k}\cdot\hat{\mathbf{n}}) \,.
\end{equation}
Note again the appearance of the sign-function.

Finally, the appropriate normalisation constant can be determined
upon inserting the
kernel~(\ref{PropKernel}--\ref{PropKernelconst}) into the
functional Schr\"odinger equation:
\begin{equation}\label{funcschrodingereqn}
\left(i\hbar\partial_{t}-\int d^{3}\mathbf{x}
\frac{-\hbar^{2}}{2a^{3}(t)}\frac{\delta^{2}}
{\delta\phi^{2}(\mathbf{x})}+\frac{a(t)}{2}
(\overrightarrow{\nabla}\phi(\mathbf{x}))^{2}
+\frac{a^{3}(t)}{2}\left(m^{2} + \xi R\right)\phi^{2}(\mathbf{x})
\right)K(\phi,t;\phi',t') = 0\,.
\end{equation}
We consider the following general form for the kernel~(\ref{PropKernel})
in position space:
\begin{equation}\label{kernelpositionspace}
K(\phi,t;\phi',t') = \mathcal{M}(t,t')\,\exp\Bigg[ - \int
d^{3}\mathbf{y} d^{3}\mathbf{z}\,\Big\{
\phi(\mathbf{y})\phi(\mathbf{z})B(y,z)+
\phi'(\mathbf{y})\phi'(\mathbf{z})C(y,z)+
\phi(\mathbf{y})\phi'(\mathbf{z}) D(y,z)\Big\}\Bigg] \,,
\end{equation}
where we define for example $B(\mathbf{y},t;\mathbf{z},t')\equiv
B(y,z)$ for brevity and where:
\begin{equation}\label{Bfunctionfouriertransf}
B(y,z) = \int \frac{d^{3}\mathbf{k}}{(2\pi)^{3}} B(k,t,t')
e^{i\mathbf{k}\cdot(\mathbf{y}-\mathbf{z})}\,,
\end{equation}
and similarly for $C(y,z)$ and $D(y,z)$. Using
(\ref{PropKernelconst}), we can easily check that the Fourier
transform of (\ref{kernelpositionspace}) satisfies the functional
Schr\"odinger equation at each (non-zero) order in the fields
$\phi$ and $\phi'$. At zeroth order, the functional Schr\"odinger
equation reads:
\begin{equation}
i\hbar \frac{\partial}{\partial t}\mathrm{log}\mathcal{M}(t,t')
-\frac{\hbar^{2}}{a^{3}(t)} \int d^{3}\mathbf{x} B(x,x) = 0
\label{KernelposM} \,.
\end{equation}
After switching to Fourier space, we write
$\mathcal{M}(t,t')=\prod_{\mathbf{k}}
\mathcal{M}_{\mathbf{k}}(t,t')$. Next, it is a non-trivial step to
exploit the functional Schr\"odinger equation at order $\phi\phi'$
in order to explicitly preserve the invariance under
$\mathbf{k}\rightarrow -\mathbf{k}$. This yields:
\begin{equation}\label{NormalisationM1}
i\hbar \frac{\partial}{\partial t}\log \mathcal{M}_{k}(t,t')
-\frac{i\hbar}{2} \frac{\partial}{\partial t}\log D(k,t,t') = 0\,.
\end{equation}
We can solve this differential equation straightforwardly:
\begin{equation}\label{Normalisationsolution}
\mathcal{M}_{k}(t,t') = \mathcal{M}_{0,k}\sqrt{D(k,t,t')}\,,
\end{equation}
where the time independent constant $\mathcal{M}_{0,k}$ has to be
fixed by condition (\ref{symreqK2}), or, alternatively by
requiring $\int\mathcal{D}\phi K(\phi,t;\phi',t) = 1$. The result
is: $\mathcal{M}_{0, k}=1/\sqrt{-2\pi V}$. Hence, the
normalisation constant is given by:
\begin{equation}\label{PropKernelconstM}
\mathcal{M}_{k}(t,t') = \sqrt{\frac{-D(k,t,t')}{2\pi V}} \,.
\end{equation}
Finally, we can straightforwardly check by substitution that
symmetry requirement (\ref{symreqK1}) is also met. The kernel we
have just constructed can readily be applied to many applications
when dealing with non-interacting scalar fields in homogeneous
spacetimes, as we will come to discuss in section
\ref{Applications}.

\section{The Kernel in the presence of a source}
\label{The Kernel in the Interacting Case}

So far we have only examined the kernel for a free,
non-interacting, quantum field. It is thus natural to turn to the
interacting case which is particularly interesting for many
physical situations. We will generalise the action in equation
(\ref{action}) and incorporate a spacetime dependent source term
$J(x)$ coupled linearly to $\phi(x)$. This particular type of
interaction can model interactions with other scalar or fermionic
fields, provided it is linear in one of the scalar fields. The
action is given by:
\begin{equation} \label{actionS}
S[\phi] = \int d^{4}x \, \sqrt{-g} \left( -\frac{1}{2}\,
\partial_{\alpha} \phi(x) \, \partial_{\beta} \phi(x) \, g^{\alpha\beta} -
\frac{1}{2} \,\left( m^{2} +\xi R\right) \phi^{2}(x) +J(x)\phi(x)
\right )\,,
\end{equation}
where, again, the time integral is performed over the finite
interval from $t'$ to $t$ and the position integral is over all
space. This action leads to the following equation of motion for
the classical field:
\begin{equation} \label{motionphiS}
\Box \phi_{\mathrm{cl}}(x) -\left( m^{2} +\xi R\right)
\phi_{\mathrm{cl}}(x)=-J(x)\,.
\end{equation}
Fourier transforming and using the FLRW metric as before yields:
\begin{equation} \label{motionphi3S}
\left ( \frac{\partial^{2}}{\partial {t}^{2}}+3H\frac{\partial}
{\partial t}+\frac{k^{2}} {a^{2}(t)}+  m^{2} +\xi
R\right)\phi_{\mathbf{k}}(t)=J_{\mathbf{k}}(t)\,,
\end{equation}
analogously to equation (\ref{motionphi2}). We split the quantum
field as in~(\ref{qfielcfield}) and require that the boundary
conditions in the path integral in
(\ref{projection-functionalintegral}) are carried solely by the
classical field. We thus arrive at:
\begin{equation}\label{actionrewrite1S}
S[\phi_{\mathrm{cl}}+\delta\phi] = S_{0}[\delta\phi] + \frac{1}{2}
\int d^{3}\mathbf{x}\, \phi_{\mathrm{cl}}(x) \,
\pi_{\mathrm{cl}}(x) \Big|_{t'}^{t} + \frac{1}{2} \int d^{4}x
\sqrt{-g}\ \phi_{\mathrm{cl}}(x) J(x)\,.
\end{equation}
First of all note that $S_{0}[\delta\phi]$ refers to the
contribution to the action of $\delta\phi$ in the absence of
interactions. Secondly, when comparing to the non-interacting case,
we see that the second term in the equation above
is unchanged and the source enters only through the third term.
Analogous to equation (\ref{actionrewrite2}) the equation above in
Fourier space is given by:
\begin{equation}\label{actionrewrite2S}
S[\phi_{\mathrm{cl}}+\delta\phi] = S_{0}[\delta\phi] +
\left.\frac{1}{4} \, \int \frac{d^{3}\mathbf{k}}{(2\pi)^{3}}\,
\left[ a^{3}(\tilde{t}\,)
\partial_{\tilde{t}}\left | \phi_{\mathbf{k}} (\tilde{t}\,)\right |^{2}\right]
\right|_{\tilde{t}=t'}^{\tilde{t}=t} + \frac{1}{2}\int
\frac{d^{3}\mathbf{k}}{(2\pi)^{3}}\int_{t'}^{t}d\tilde{t}\,a^{3}(\tilde{t}\,)
J_{\mathbf{k}}^{\ast}(\tilde{t}\,)\,
\phi_{\mathbf{k}}(\tilde{t}\,) \,.
\end{equation}
The homogeneous solution of (\ref{motionphi3S}) is given by
(\ref{phisolutionpropagator}). Employing the Green's function
method and noting that only the homogeneous solution carries the
boundary conditions, we find the following total solution:
\begin{equation}\label{solutionlincombS1}
\phi_{\mathbf{k}}(\tilde{t}\,)= \phi_{\mathbf{k}}
\frac{G_{\mathbf{k}}(\tilde{t},t')}{G_{\mathbf{k}}(t,t')} +
\phi'_{\mathbf{k}}
\frac{G_{\mathbf{k}}(t,\tilde{t}\,)}{G_{\mathbf{k}}(t,t')} +
S_{\mathbf{k}}(\tilde{t}\,) \,,
\end{equation}
where:
\begin{equation}\label{Sourcesolution1}
S_{\mathbf{k}}(\tilde{t}) = \int_{t'}^{t} d\tau\,\,
Y_{\mathbf{k}}(\tilde{t},\tau) \,J_{\mathbf{k}}(\tau)\,,
\end{equation}
and where $Y_{\mathbf{k}}(\tilde{t},\tau)$ is the appropriate
Green's function corresponding to (\ref{motionphi3S}):
\begin{equation}\label{Sourcesolution}
Y_{\mathbf{k}}(\tilde{t},\tau) = -\theta(\tilde{t}-\tau)
\frac{G_{\mathbf{k}}(\tilde{t},\tau)}{W_{\mathbf{k}}(\tau)}
+\frac{G_{\mathbf{k}}(\tilde{t},t') G_{\mathbf{k}}(t,\tau)}
{G_{\mathbf{k}}(t,t')W_{\mathbf{k}}(\tau)} \,.
\end{equation}
Finally, we easily derive the symmetry relation
$S_{-\mathbf{k}}(t)=S^{\ast}_{\mathbf{k}}(t)$. Upon recalling the
definition of the Wronskian, $W_{\mathbf{k}}(t) =
\left[\partial_{t'} G_{\mathbf{k}}(t,t')\right]_{t'\rightarrow
t}$, we see that the Green's function~(\ref{Sourcesolution}) is,
as expected, expressed solely in terms of the causal propagator.
We proceed completely analogously by substituting expansion
(\ref{solutionlincombS1}) into the action, equation
(\ref{actionrewrite2S}), collecting all terms at each order, i.e.:
$\phi_{\mathbf{k}}\phi_{\mathbf{k}}^{\ast}$,
$\phi'_{\mathbf{k}}{\phi'}_{\mathbf{k}}^{\ast}$,
$\phi_{\mathbf{k}}{\phi'}_{\mathbf{k}}^{\ast}$ and
$\phi^{\ast}_{\mathbf{k}}{\phi'}_{\mathbf{k}}$ at quadratic order
and $\phi_{\mathbf{k}}$, $\phi^{\ast}_{\mathbf{k}}$,
$\phi'_{\mathbf{k}}$ and ${\phi'}^{\ast}_{\mathbf{k}}$ as the
linear contributions. The source contributes through the linear
terms only and the quadratic ones remain unaffected. Furthermore,
we can safely omit the terms at zeroth order in the fields,
because we will incorporate those into the overall normalisation.
Hence the full result for the kernel reads:
\begin{eqnarray}\label{PropKernelS}
K(\phi,t;\phi',t') &=&
\prod_{\mathbf{k}}\,\mathcal{M}_{\mathbf{k}}\, \exp\Bigg[
-\frac{1}{V}\Big\{B(k,t,t')\phi_{\mathbf{k}}\phi^{\ast}_{\mathbf{k}}
+ C(k,t,t')\phi'_{\mathbf{k}}{\phi'}^{\ast}_{\mathbf{k}}
 + \frac{1}{2}D(k,t,t')\Big(\phi_{\mathbf{k}}
{\phi'}^{\ast}_{\mathbf{k}} +
\phi^{\ast}_{\mathbf{k}}{\phi'}_{\mathbf{k}}\Big)
\\
&& \qquad\qquad\qquad +
\frac{1}{2}E(\mathbf{k},t,t')\phi_{\mathbf{k}}-
\frac{1}{2}E^{\ast}(\mathbf{k},t,t')\phi^{\ast}_{\mathbf{k}}
+\frac{1}{2}F(\mathbf{k},t,t')\phi'_{\mathbf{k}}-
\frac{1}{2}F^{\ast}(\mathbf{k},t,t'){\phi'}^{\ast}_{\mathbf{k}}
\Big \} \Bigg] \nonumber\,,
\end{eqnarray}
where $\mathcal{M}_{\mathbf{k}}$ is again a normalisation constant
and where the $B$-, $C$- and $D$-functions are given by
(\ref{PropKernelconst}) and, finally, where:
\begin{subequations}
\begin{eqnarray}
E(\mathbf{k},t,t') &=& -\frac{i}{2\hbar}\Bigg[
a^{3}(t)\partial_{t} S^{\ast}_{\mathbf{k}}(t) +
\int_{t'}^{t}d\tilde{t} \, a^{3}(\tilde{t}\,)
J^{\ast}_{\mathbf{k}}(\tilde{t}\,)
\frac{G_{\mathbf{k}}(\tilde{t},t')} {G_{\mathbf{k}}(t,t')}\Bigg]
\label{PropKernelconstSE} \\
F(\mathbf{k},t,t') &=& \frac{i}{2\hbar}\Bigg[
a^{3}(t')\partial_{t'} S^{\ast}_{\mathbf{k}}(t')-
\int_{t'}^{t}d\tilde{t} \, a^{3}(\tilde{t}\,)
J^{\ast}_{\mathbf{k}}(\tilde{t}\,)
\frac{G_{\mathbf{k}}(t,\tilde{t})} {G_{\mathbf{k}}(t,t')} \Bigg]
\label{PropKernelconstSF} \,.
\end{eqnarray}
\end{subequations}
For a detailed derivation see \cite{Koksma:2007zz}. The derivative
$\partial_{t} S_{\mathbf{k}}(t)$ should be interpreted as
$\partial_{\tilde{t}\,} S_{\mathbf{k}}
(\tilde{t}\,)|_{\tilde{t}=t}$. Since $E^{\ast}(\mathbf{k},t,t')=
-E(-\mathbf{k},t,t')$, note that the contribution at order
$\phi^{\ast}_{\mathbf{k}}$ carries the opposite sign as compared
to the contribution at order $\phi_{\mathbf{k}}$ which {\em
mutatis mutandis} holds for $F(\mathbf{k},t,t')$ and
$F^{\ast}(\mathbf{k},t,t')$.

Let us now find the new normalisation constant. The functional
Schr\"odinger equation changes in comparison with
(\ref{funcschrodingereqn}) to:
\begin{equation}\label{funcschrodingereqnS}
\left(i\hbar\partial_{t}-\int d^{3}\mathbf{x}
\frac{-\hbar^{2}}{2a^{3}(t)}\frac{\delta^{2}}
{\delta\phi^{2}(\mathbf{x})}+\frac{a(t)}{2}
(\overrightarrow{\nabla}\phi(\mathbf{x}))^{2}
+a^{3}(t)\left\{\frac{1}{2}\left(m^{2}+\xi R\right)
\phi^{2}(\mathbf{x}) -J(\mathbf{x})\phi(\mathbf{x})\right\}
\right) K(\phi,t;\phi',t') = 0.
\end{equation}
The position space form for the kernel~(\ref{PropKernelS})
generalises equation (\ref{kernelpositionspace}) to:
\begin{eqnarray}\label{kernelpositionspaceS}
K(\phi,t;\phi',t') &=& \mathcal{M}(t,t')\,\exp\Bigg[ - \int
d^{3}\mathbf{y} d^{3}\mathbf{z}\,\Big\{
\phi(\mathbf{y})\phi(\mathbf{z})B(y,z)+
\phi'(\mathbf{y})\phi'(\mathbf{z})C(y,z) +
\phi(\mathbf{y})\phi'(\mathbf{z})
D(y,z)\Big\} \nonumber \\
&& \qquad\qquad\qquad - \int d^{3}\mathbf{y} \Big\{
\phi(\mathbf{y})E(\mathbf{y},t,t') +
\phi'(\mathbf{y})F(\mathbf{y},t,t')\Big\}\Bigg]\,.
\end{eqnarray}
Substitution into the functional Schr\"odinger equation
(\ref{funcschrodingereqnS}) leads again to a number of equations
at various orders in the fields $\phi$ and $\phi'$ that are indeed
satisfied simultaneously. Writing $\mathcal{M}(t,t')=
\prod_{\mathbf{k}} \mathcal{M}_{\mathbf{k}}(t,t')$, the zeroth
order equation in Fourier space reads:
\begin{equation}\label{NormalisationMS1}
i\hbar \frac{\partial}{\partial t}\mathrm{log}
\mathcal{M}_{\mathbf{k}}(t,t') -\frac{\hbar^{2}}{a^{3}(t)}\Big\{
B(k,t,t')-\frac{1}{2V}|E(\mathbf{k},t,t')|^{2}\Big\} = 0\,.
\end{equation}
We can solve this equation straightforwardly by:
\begin{equation}\label{NormalisationsolutionS}
\mathcal{M}_{\mathbf{k}}(t,t') =
\mathcal{M}_{0,\mathbf{k}}\sqrt{D(k,t,t')}
\exp\left[\frac{i\hbar}{2V}\int_{t'}^{t}d\tilde{t}\,
\frac{|E(\mathbf{k},\tilde{t},t')|^{2}}{a^{3}(\tilde{t}\,)} \right
] \,,
\end{equation}
where $\mathcal{M}_{0,\mathbf{k}}$ is again a time independent
constant. Note that the lower boundary of the integral is $t'$
because our solution for the kernel is constructed such that it
vanishes for times less than $t'$.

Finally, $\mathcal{M}_{0,\mathbf{k}}$ has to be fixed by condition
(\ref{symreqK2}) as before. Now, in the limit when $\Delta t =
t-t'\,\rightarrow 0$ the kernel can easily be verified to be:
\begin{eqnarray}\label{PropKernellimitS}
K(\phi,t;\phi',t) &=&
\prod_{\mathbf{k}}\,\mathcal{M}_{0,\mathbf{k}}\sqrt{
\frac{ia^{3}(t)} {\hbar\Delta t}}
\exp\Bigg[-\frac{a^{3}(t)}{2i\hbar V \Delta t}
\Big(\phi_{\mathbf{k}}-\phi'_{\mathbf{k}} -\frac{\partial_{t}
S_{\mathbf{k}}(t)}{2}\Delta t \Big)\Big(\phi^{\ast}_{\mathbf{k}} -
{\phi'}^{\ast}_{\mathbf{k}} + \frac{\partial_{t}
S^{\ast}_{\mathbf{k}}(t)}{2}\Delta t \Big)  \\
&& \qquad\qquad\qquad \qquad\qquad+\frac{ia^{3}(t)}{8\hbar
V}\partial_{t} S^{\ast}_{\mathbf{k}}(t)\partial_{t}
S_{\mathbf{k}}(t) \Bigg] \nonumber\,.
\end{eqnarray}
The dominant terms in the exponent behave as $1/\Delta t$ as
$\Delta t \rightarrow 0$ and, consequently, the terms involving
the source all give a vanishing contribution in this limit.
Equivalently, the shift of the delta-function induced by the
source vanishes at the lower boundary. We conclude therefore that
(\ref{PropKernellimitS}) indeed reduces to a representation of a
delta-function correctly. The result for
$\mathcal{M}_{0,\mathbf{k}}$ is hence unchanged compared to the
free case (\ref{PropKernelconstM}). Furthermore, note that the
width of the Gaussian in the limit when $\Delta t \rightarrow 0$
behaves as $\Delta_{\phi}^{2} \propto \Delta t$, which indicates
an early time diffusive wave packet spreading in configuration
space. Finally, it can easily be verified that the obtained
solution satisfies the correct symmetry properties
(\ref{symreqK1}).

\section{Causality in Quantum Mechanics}
\label{Causality in Quantum Mechanics}

Let us now examine the causal structure of quantum field theory in
the formalism developed above. In this section we will prove that
a) the causal propagator (\ref{commutatorpropagator}) in position
space vanishes when $\|\mathbf{x}-\mathbf{x}'\|>|\eta-\eta'|$ in
de Sitter space and b) the same light cone structure is also
present in the Functional Schr\"odinger picture and can be
inferred from the kernel we have constructed. Here $\eta$ denotes
conformal time defined by $d\eta = dt/a(t)$. The clear geometrical
interpretation of causality in position space cannot be easily
transferred to Fourier space because a Fourier transformation is
nonlocal: one cannot simply draw ``light cones'' in Fourier space.
A delicate cancellation occurs such that a superposition of
Fourier modes cancel precisely outside past and future light
cones.

Since FLRW spacetimes are conformal, their causality structure is
most easily described in conformal coordinates, in which the light
cones are simply $\|\mathbf{x}-\mathbf{x}'\|=|\eta-\eta'|$. In de
Sitter space the scale factor $a(t)={\rm e}^{Ht}$ implies
$a(\eta)=-1/(H\eta)$, $\eta<0$ and $H$ is the Hubble parameter of
de Sitter space. In de Sitter space in $D$ dimensions the
Chernikov-Tagirov~\cite{Chernikov:1968zm} scalar propagator
reads~\cite{Garbrecht:2006jm,Prokopec:2003tm}:
\begin{eqnarray}\label{scalarpropagator}
i\Delta_{ab}(x,x') &=& \frac{H^{D-2}}{(4\pi)^{D/2}}
\Gamma\left(\frac{D}{2}-1\right)\frac{1}{y_{ab}^{\frac{D}{2}-1}} \\
&& + \frac{H^{2}}{16\pi^{2}}\sum_{n=0}^{\infty} \frac{
\Gamma\left(\frac{3}{2}+\nu+n\right)
\Gamma\left(\frac{3}{2}-\nu+n\right)}
{\Gamma\left(\frac{1}{2}+\nu\right)
\Gamma\left(\frac{1}{2}-\nu\right)}
\left(\frac{y_{ab}}{4}\right)^{n}\Big[ \ln\left(\frac{y_{ab}}
{4}\right) + \psi\left(\frac{3}{2}+\nu+n\right) +
\psi\left(\frac{3}{2}-\nu+n\right) \nonumber
\\
&&\qquad\qquad\qquad\qquad\qquad\qquad\qquad\qquad\qquad\qquad
-\psi\left(1+n\right) -
\psi\left(2+n\right)\Big]+\mathcal{O}(D-4)\nonumber \,,
\end{eqnarray}
where we have kept the $D$-dimensional form of the first (most
singular) term that may lead to singular contributions in the
limit when $D\rightarrow 4$. Note that $a$ and $b$ can be either
$+$ or $-$, $\nu^{2}=9/4-(m^{2}+\xi R)/H^{2}$, $R=12H^2$ and
$\psi(z)=d[\ln\Gamma(z)]/dz$ as usual. Furthermore, $y_{ab}$ is
given by:
\begin{equation}\label{yfunction}
y_{ab} = aa'H^{2}\Delta x^{2}_{ab} \,.
\end{equation}
The scalar function $y={y_{ab}}|_{\epsilon\rightarrow 0}
  =(1/4) \sin^2(H\ell/2)$ is a simple function of the geodesic distance
 $\ell$ in de Sitter space.
Since we are only interested in the two Wightman functions
contributing to the causal propagator, we have:
\begin{eqnarray}
\Delta x^{2}_{-+} &=& -(\eta - \eta'-i\epsilon)^{2}+ r^{2}
\label{xseparation1} \\
\Delta x^{2}_{+-} &=& -(\eta - \eta'+i\epsilon)^{2}+ r^{2}
\label{xseparation2} \,,
\end{eqnarray}
where $r=\|\mathbf{x}-\mathbf{x}'\|$. The causal propagator
(\ref{commutatorpropagator}) thus follows as:
\begin{eqnarray}\label{scalarpropagator2}
\left\langle \Omega\left|\left[\hat{\phi}(x),
\hat{\phi}(x')\right]\right|\Omega\right\rangle &=&
i\Delta_{-+}(x,x') - i\Delta_{+-}(x,x') =
\frac{H^{D-2}}{(4\pi)^{D/2}}
\Gamma\left(\frac{D}{2}-1\right)\left\{\frac{1}{y_{-+}^{\frac{D}{2}-1}}
- \frac{1}{y_{+-}^{\frac{D}{2}-1}}\right\}  \\
&& + \frac{H^{2}}{16\pi^{2}}\sum_{n=0}^{\infty} \frac{
\Gamma\left(\frac{3}{2}+\nu+n\right)
\Gamma\left(\frac{3}{2}-\nu+n\right)}
{\Gamma\left(\frac{1}{2}+\nu\right)
\Gamma\left(\frac{1}{2}-\nu\right)}
\left(\frac{y_{}}{4}\right)^{n}\left\{  \ln\left(y_{-+}\right) -
\ln\left(y_{+-}\right) \right\}\nonumber \,.
\end{eqnarray}
Clearly, the first term represents the standard Hadamard
singularity that every propagator contains. The two terms relevant
to our calculation are the ones in curly parentheses. It can be
shown that:
\begin{eqnarray}\label{scalarpropagator3}
\frac{1}{y_{-+}^{\frac{D}{2}-1}} -
\frac{1}{y_{+-}^{\frac{D}{2}-1}} &=& \frac{1}{aa'H^{2}} \Big[
-2\pi i\,\,\mathrm{sgn}(\Delta\eta)
\delta(r^{2}-\Delta\eta^{2})\Big]
 + {\cal O}(D-4) \\
\ln\left(y_{-+}\right) - \ln\left(y_{+-}\right) &=& 2\pi i
[\theta(\Delta\eta-r)-\theta(-\Delta\eta-r)] \,,
\label{scalarpropagator3:ln}
\end{eqnarray}
where $\Delta\eta=\eta-\eta'$. While the Hadamard pole results in
a singular contribution at the light cone $|\Delta\eta|=r$, the
logarithmic cuts -- which are a mathematical description for
amplification of super-Hubble correlations occurring in
accelerating spacetimes -- are responsible for finite
contributions within past and future light cones. Outside past and
future light cones both
contributions~(\ref{scalarpropagator3}--\ref{scalarpropagator3:ln})
vanish, as expected. Therefore, our causal propagator
(\ref{scalarpropagator2}) also vanishes in these regions as it
should. Although we performed the calculation in de Sitter
spacetime, an analogous result should hold for more general
spacetimes.

Now, we will show that our kernel analysis is consistent with the
arguments put forward above. In the Functional Schr\"odinger
picture the causal propagator reads:
\begin{eqnarray}\label{vacuumexpvalufuncschrodinger}
&& \left\langle \Omega\left|\left[\hat{\phi}(\mathbf{x},t_{1}),
\hat{\phi}(\mathbf{x}',t_{2})\right]\right|\Omega\right\rangle \\
&&\qquad\qquad= \int \mathcal{D}\phi\mathcal{D}\phi'
\phi(\mathbf{x}) \phi(\mathbf{x}')\left\{ \Psi^{\ast}(\phi,t_{1})
\Psi(\phi',t_{2}) K(\phi,t_{1};\phi',t_{2}) - \Psi(\phi,t_{1})
\Psi^{\ast}(\phi',t_{2})
K(\phi',t_{2};\phi,t_{1})\right\}\nonumber\,.
\end{eqnarray}
For simplicity, we consider the free kernel (\ref{PropKernel}).
The Gaussian vacuum wave functional (see
\cite{Guth:1985ya,Prokopec:2006fc,Long:1996wf}) is given by:
\begin{equation} \label{vacuumwavefunctional}
\Psi(\phi, t)=\mathcal{N}(t)\,\,\exp\left [-\int d^{3}\mathbf{y}
d^{3}\mathbf{z}\,
\phi(\mathbf{y})A(\mathbf{y},\mathbf{z},t)\phi(\mathbf{z})\right ]
= \prod_{\mathbf{k}} \mathcal{N}_{\mathbf{k}}(t)\,\,\exp\left
[-\frac{1}{V}\Big\{\phi_{\mathbf{k}}^{\ast}\,A_{k}(t)\,\phi_{\mathbf{k}}\Big\}
\right ] \,,
\end{equation}
where $\mathcal{N}(t)$ is a normalisation constant formally given
by:
\begin{equation} \label{normconstn}
\mathcal{N}(t)\,= \mathcal{N}_{0}\,\,\exp\left [ -i\hbar \int
dt\int d^{3}\mathbf{x}\frac{A(\mathbf{x},\mathbf{x},t)}{a^{3}(t)}
\right ]\,,
\end{equation}
and where the $A$-function in Fourier space is given by:
\begin{equation} \label{vacuumwavefunctional2}
A_{k}(t)=\frac{1}{2i\hbar}\,a^{3}(t) \frac{\partial}{\partial
t}\Big[\mathrm{log}\Big\{\theta(-\mathbf{k}\cdot\hat{\mathbf{n}})
\phi_{\mathbf{k}}(t) + \theta(\mathbf{k}\cdot\hat{\mathbf{n}})
\phi_{\mathbf{k}}^{\ast}(t)\Big\} \Big]\,,
\end{equation}
where $k=\|\mathbf{k}\|$ and where $\hat{\mathbf{n}}$ is a unit
vector normal to an arbitrary plane through the origin in
$\mathbf{k}$-space. We see after substitution in the functional
Schr\"odinger equation that $\phi_{\mathbf{k}}(t)$ and
$\phi_{\mathbf{k}}^{\ast}(t)$ in fact obey (\ref{motionphi2}).
This solution differs slightly when compared to~\cite{Guth:1985ya}.
The form of~(\ref{vacuumwavefunctional2})
is dictated by the required invariance of $A_{k}(t)$ under
$\mathbf{k}\rightarrow -\mathbf{k}$, and
the difference is due to our definition of Fourier decomposition
in terms of complex mode functions~(\ref{fieldfouriertransform}).

In principle, we could allow for an even more general argument of
the logarithm, namely
$a_k[\theta(-\mathbf{k}\cdot\hat{\mathbf{n}}) \phi_{\mathbf{k}}(t)
+ \theta(\mathbf{k}\cdot\hat{\mathbf{n}})
\phi_{\mathbf{k}}^{\ast}(t)] + b_k
[\theta(\mathbf{k}\cdot\hat{\mathbf{n}}) \phi_{\mathbf{k}}(t) +
\theta(-\mathbf{k}\cdot\hat{\mathbf{n}})
\phi_{\mathbf{k}}^{\ast}(t)]$. Then, we require
$|a_k|^{2}-|b_k|^{2}=1$ and in order to preserve homogeneity
$\mathrm{arg}(a_k)=\mathrm{arg}(b_k)$. These states would
correspond to the most general pure (minimum uncertainty) states
with a ``non-zero particle number'' $|b_k|^{2}$.

Splitting $A_{k}(t)$ into real and imaginary parts yields:
\begin{equation} \label{vacuumwavefunctional3}
A_{k}(t)= \frac{1}{4\hbar|\phi_{\mathbf{k}}(t)|^{2}}\left(1 - i
a^{3}(t)\frac{\partial}{\partial
t}\,|\phi_{\mathbf{k}}(t)|^{2}\right)\,.
\end{equation}
We can in fact perform the two functional integrals in
(\ref{vacuumexpvalufuncschrodinger}) by introducing two additional
sources in the exponent, varying with respect to these sources and
setting them to zero subsequently:
\begin{eqnarray}\label{vacuumexpvalufuncschrodinger2}
&& \left\langle \Omega\left|\left[\hat{\phi}(\mathbf{x},t_{1}),
\hat{\phi}(\mathbf{x}',t_{2})\right]\right|\Omega\right\rangle \\
&&\quad= \left.\frac{\delta}{\delta J(\mathbf{x})}
\frac{\delta}{\delta J'(\mathbf{x}')}  \int
\mathcal{D}\phi\mathcal{D}\phi'  \left\{ \Psi^{\ast}(\phi,t_{1})
\Psi(\phi',t_{2}) K_{J,J'}(\phi,t_{1};\phi',t_{2}) -
\Psi(\phi,t_{1}) \Psi^{\ast}(\phi',t_{2})
K_{J,J'}(\phi',t_{2};\phi,t_{1})\right\}\right|_{J=J'=0}
\nonumber,
\end{eqnarray}
where:
\begin{equation}\label{vacuumexpvalufuncschrodinger2b}
K_{J,J'}(\phi,t_{1};\phi',t_{2}) = K(\phi,t_{1};\phi',t_{2})
\exp\left[-\int
d^{3}\mathbf{z}\{J(\mathbf{z})\phi(\mathbf{z})+J'(\mathbf{z})\phi'(\mathbf{z})
\}\right]\,.
\end{equation}
After some algebra we arrive at the following intermediate result,
where the two Wightman functions can clearly be recognised:
\begin{eqnarray}\label{vacuumexpvalufuncschrodinger3}
\left\langle \Omega\left|\left[\hat{\phi}(\mathbf{x},t_{1}),
\hat{\phi}(\mathbf{x}',t_{2})\right]\right|\Omega\right\rangle
&=& -\int\frac{d^{3}\mathbf{k}}{(2\pi)^{3}}
\left\{\frac{D(k,t_{1},t_{2})} {4(C(k,t_{1},t_{2})+A_{k}(t_{2}))
(B(k,t_{1},t_{2})+A_{k}^{\ast}(t_{1})) -D^{2}(k,t_{1},t_{2})} \right.\\
&& + \left.\frac{D(k,t_{1},t_{2})}
{4(C(k,t_{1},t_{2})-A_{k}^{\ast}(t_{2}))
(B(k,t_{1},t_{2})-A_{k}(t_{1})) -D^{2}(k,t_{1},t_{2})}\right\}
e^{i\mathbf{k}\cdot(\mathbf{x}-\mathbf{x}')}\nonumber\,.
\end{eqnarray}
Note that both $\Delta_{-+}$ and $\Delta_{+-}$ are expressed in
terms of causal Green's functions and the initial state through
$A_{k}(t)$. The Functional Schr\"odinger picture shows that the
evolution of correlators can be fully expressed in terms of causal
propagators and initial state only. This is to be contrasted to
out-of-equilibrium field theories in Heisenberg picture, where
often one reads that in non-equilibrium problems there are two
independent two point functions: the spectral function (causal
propagator, commutator) and the statistical propagator
(anticommutator).\footnote{This comment is formal in nature and in
fact we do not question the utility of considering separately the
evolution of statistical and causal
propagators~\cite{Berges:2004yj}.} Employing the definitions of
the occurring functions in terms of the fundamental solutions in
equations (\ref{PropKernelconst}) and
(\ref{vacuumwavefunctional2}), allows us to even further simplify
this result to obtain:
\begin{eqnarray}\label{vacuumexpvalufuncschrodinger4}
\left\langle \Omega\left|\left[\hat{\phi}(\mathbf{x},t_{1}),
\hat{\phi}(\mathbf{x}',t_{2})\right]\right|\Omega\right\rangle
&=& \hbar \int\frac{d^{3}\mathbf{k}}{(2\pi)^{3}}
G_{\mathbf{k}}(t_{1},t_{2})\mathrm{sgn}(\mathbf{k}\cdot\hat{\mathbf{n}})
e^{i\mathbf{k}\cdot(\mathbf{x}-\mathbf{x}')} \\
&=& \hbar \int\frac{d^{3}\mathbf{k}}{(2\pi)^{3}}
G_{c}(k,t_{1},t_{2})e^{i\mathbf{k}\cdot(\mathbf{x}-\mathbf{x}')}
\nonumber \,,
\end{eqnarray}
as it should be. We have used identity (\ref{causalproplinkGprop}),
which establishes the link with the Heisenberg picture fields.
Note that the Wightman functions in
(\ref{vacuumexpvalufuncschrodinger3}) depend on the initial state
through $A_{k}(t)$ whereas the causal propagator does not. Hence,
we have shown that the kernel although developed in Fourier space,
preserves the causal structure in terms of light cones in position
space.

Concluding, since the commutator is causal and the kernel is
expressed solely in terms of this causal quantity, quantum
mechanics in de Sitter spacetime is fully causal. This is in
contrast to claims made, for example, in \cite{Nomoto:1992cp,
Hegerfeldt:1974qu, Hegerfeldt:1985fy, Hegerfeldt:1998ar,
Halvorson:2006wj, Halvorson:2001hb, FernandoPerez:1976ib}. Apart
from \cite{Nomoto:1992cp}, the claims in the
literature~\cite{Hegerfeldt:1974qu, Hegerfeldt:1985fy,
Hegerfeldt:1998ar, Halvorson:2006wj, Halvorson:2001hb,
FernandoPerez:1976ib} are based on considering single particle
quantum mechanics, in contrast to the Functional Schr\"odinger
picture used in the present work. This means that causality of
quantum mechanics can be fully appreciated only within the context
of (relativistic) quantum field theory, and can lead to misleading
results when viewed within the one particle formulation of quantum
mechanics.

If initial and final times are equal, the
kernel~(\ref{PropKernel}--\ref{PropKernelconst}) reduces to a
delta-function, i.e.: there is no propagation of the field. If
both times differ, propagation is dictated by the commutator
propagator and hence causality is preserved in general in quantum
mechanics. Note finally that equations (\ref{expectationvalueQ2})
and (\ref{expectationvalueQ3}) support this statement. Although we
have proved this consistency in de Sitter spacetime explicitly, we
have no reason to believe that our result does not hold in FLRW
and more general spacetimes. In fact, one could take the region in
spacetime where the commutator
(\ref{vacuumexpvalufuncschrodinger4}) vanishes as the definition
for light cones in general spacetimes.
\begin{figure}[htbp]
 \centering
  \begin{minipage}[t]{.65\textwidth}
   \includegraphics[width=\textwidth]{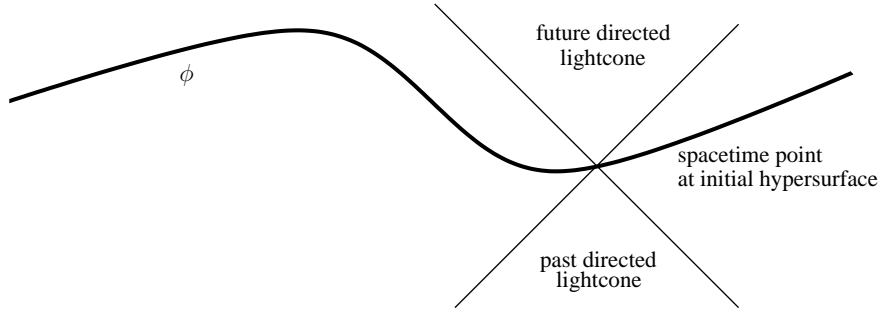}
   {\em \caption{Causality in quantum mechanics. The kernel is expressed in
   terms of the causal commutator exclusively. This ensures that a certain spacetime point
   can only affect spacetime points in its future directed light cone. Likewise, a certain
   spacetime point can only be affected by other spacetime points
   in its past directed light cone.
   \label{fig:lightcone} }}
   \end{minipage}
\end{figure}

\section{Applications}
\label{Applications}

To illustrate the applicability of the kernel, we will turn our
attention to three explicit examples. We construct the kernel in
both Minkowski and pure de Sitter spacetimes, where in the former
case one can recognise the simple harmonic oscillator kernel.
Finally, we will examine the evolution of the vacuum state for a
non-interacting Hamiltonian in general FLRW spacetimes.

\subsection{Example I: Simple Harmonic Oscillator} \label{Example
I: Simple Harmonic Oscillator}

Consider the simple harmonic oscillator toy model. Let us take the
Minkowski metric, i.e.: we set $a(t)=1$ in the FLRW metric. In the
minimally coupled case ($\xi=0$), the equation of motion that the
field modes $\phi_{\mathbf{k}}(t)$ obey follows from
(\ref{motionphi2}) as:
\begin{equation}\label{SHOeqnofmotion}
\left(\partial_{t}^{2}+\omega^{2}\right) \phi_{\mathbf{k}}(t)=0\,,
\end{equation}
where $\omega^{2}\equiv k^{2}+ m^{2}$. If we choose our
fundamental solution of this equation of motion to be:
\begin{equation}\label{SHOsolution}
\chi_{\mathbf{k}}(t)=\frac{1}{\sqrt{2\omega}}\,\mathrm{e}^{i\omega
t}\theta(\mathbf{k}\cdot\hat{\mathbf{n}}) +
\frac{1}{\sqrt{2\omega}}\,\mathrm{e}^{-i\omega
t}\theta(-\mathbf{k}\cdot\hat{\mathbf{n}}) \,,
\end{equation}
then this solution obeys the correct symmetry properties. Note
that this choice is consistent with the Wronskian normalisation
condition (\ref{wronskian}). Again, we could have chosen an even
more general solution of the form $\vartheta_{\mathbf{k}}(t) = a_k
\chi_{\mathbf{k}}(t) + b_k \chi_{-\mathbf{k}}(t)$ and require
$|a_k|^{2}-|b_k|^{2}=1$. The Fourier transform of the causal
propagator defined in (\ref{commutatorpropagatorF}) is hence given
by:
\begin{equation}\label{SHOpropagator}
G_{\mathbf{k}}(t,t')\,= \frac{i}{\omega}\, \mathrm{sin}\left (
\omega(t-t')\right )\mathrm{sgn}(\mathbf{k}\cdot\hat{\mathbf{n}})
\,.
\end{equation}
The kernel for the simple harmonic oscillator follows as:
\begin{equation}\label{SHOpropagatorkernelfull}
K(\phi,t;\phi',t')=\prod_{\mathbf{k}}\sqrt{\frac{\omega}{2\pi
i\hbar V\mathrm{sin}(\omega(t-t'))}} \exp\left[
\frac{i\omega}{2\hbar V \mathrm{sin}(\omega(t-t'))} \Big\{
(\phi_{\mathbf{k}}\phi^{\ast}_{\mathbf{k}}+{\phi'}_{\mathbf{k}}^{\ast}
\phi'_{\mathbf{k}})
\mathrm{cos}(\omega(t-t'))\,-\,2\phi_{\mathbf{k}}{\phi'}_{\mathbf{k}}^{\ast}
\Big\}\right] \,.
\end{equation}
If we let $V\rightarrow 1$ this is indeed in agreement with
standard quantum mechanical results. It reproduces for example
\cite{Sakurai} where $m=1$.

\subsection{Example II: De Sitter Universe} \label{Example II: De
Sitter Universe}

As a second example we consider the kernel for the inflationary de
Sitter Universe \cite{Guven:1987bx}. The solution of
(\ref{motionphi2}) in conformal time is thus given by:
\begin{equation} \label{phisolutiondesitt}
\phi_{\mathbf{k}}(\eta)= \alpha_{k} \chi_{\mathbf{k}}(\eta) +
\beta_{k} \chi_{\mathbf{k}}^{\ast}(\eta)\,,
\end{equation}
where $\alpha_{k}$ and $\beta_{k}$ are two coefficients and:
\begin{equation}\label{hankelsolutions1}
\chi_{\mathbf{k}}(\eta) = \frac{1}{a(\eta)}
\sqrt{-\frac{\pi\eta}{4}}\,e^{i\frac{\pi}{2}(\nu+\frac{1}{2})}\,
H_{\nu}^{(1)}(-k\eta) \theta(\mathbf{k}\cdot\hat{\mathbf{n}}) +
\frac{1}{a(\eta)}
\sqrt{-\frac{\pi\eta}{4}}\,e^{-i\frac{\pi}{2}(\nu+\frac{1}{2})}\,
H_{\nu}^{(2)}(-k\eta) \theta(-\mathbf{k}\cdot\hat{\mathbf{n}}) \,.
\end{equation}
$H_{\nu}^{(1)}$ and $H_{\nu}^{(2)}$ are the Hankel functions of
the first and second kind, respectively, at order
$\nu^{2}=9/4-(m^{2}+\xi R)/H^{2}$, see equation (\ref{actionS}).
The causal propagator now follows as:
\begin{equation} \label{desittercausalprop}
G_{\mathbf{k}}(\eta,\eta')\,= \frac{\pi}{4H \left(a(\eta)
a(\eta')\right)^{3/2}}\left[H_{\nu}^{(1)}(-k\eta)
H_{\nu}^{(2)}(-k\eta') - H_{\nu}^{(2)}(-k\eta)
H_{\nu}^{(1)}(-k\eta') \right]
\mathrm{sgn}(\mathbf{k}\cdot\hat{\mathbf{n}}) \,.
\end{equation}
The causal propagator in turn fully determines our
kernel~(\ref{PropKernel}).

\subsection{Example III: The Evolution of the Vacuum State}
 \label{Example III: The evolution of the Vacuum State}

As a final example, we can apply the kernel to an initial vacuum
state for some non-interacting scalar field. Thus, initially, we
start with a vacuum state $|\Psi(t')\rangle$ and calculate its
forward time evolution according to standard quantum mechanical
lore (\ref{evolution}). We employ the Functional Schr\"odinger
picture as before and arrive at:
\begin{equation}\label{vacuumevolution}
\Psi(\phi,t)=\int
\mathcal{D}\phi'K(\phi,t;\phi',t')\Psi(\phi',t')\,.
\end{equation}
We can now conveniently exploit equation (\ref{PropKernel}) for
the kernel in the non-interacting case. The Gaussian vacuum wave
functional is given by (\ref{vacuumwavefunctional}). In order to
perform the functional integral in (\ref{vacuumevolution}), we
switch to Fourier space, complete the square and arrive at the
following intermediate result:
\begin{equation}\label{vacuumevolution2}
\Psi(\phi,t) = \prod_{\mathbf{k}} \mathcal{M}_{k}(t,t')
\mathcal{N}_{k}(t') \left(\frac{\pi V}{C(k,t,t')+
A_{k}(t')}\right)^{1/2} \exp\left[ -\frac{1}{V}\left\{
\phi_{\mathbf{k}}^{\ast} \left(
B(k,t,t')-\frac{1}{4}\frac{D^{2}(k,t,t')} {C(k,t,t')+
A_{k}(t')}\right)\phi_{\mathbf{k}}\right\}\right].
\end{equation}
Now we insert the definition of the causal propagator and the
Wronskian in terms of the fundamental solutions. Indeed, one can
show that the expression above reduces to the vacuum at time $t$:
\begin{equation}\label{vacuumevolution3}
B(k,t,t')-\frac{1}{4}\frac{D^{2}(k,t,t')} {C(k,t,t')+ A_{k}(t')} =
A_{k}(t) \,,
\end{equation}
and furthermore:
\begin{equation}\label{vacuumevolution4}
\mathcal{M}_{k}(t,t') \mathcal{N}_{k}(t') \left(\frac{\pi
V}{C(k,t,t')+ A_{k}(t')}\right)^{1/2} = \mathcal{N}_{k}(t)\,,
\end{equation}
as desired.

\section{Conclusion}
\label{Conclusion}

We constructed the kernel for a scalar field in FLRW spacetimes in
two cases: a free field (\ref{PropKernel}) and one coupled to a
source term (\ref{PropKernelS}). We showed that these kernels can
be expressed solely in terms of the causal propagator and
derivatives of the causal propagator. We have applied the general
formalism to three examples, the simple harmonic oscillator kernel
(\ref{SHOpropagatorkernelfull}), the kernel in de Sitter spacetime
and the evolution of the Gaussian vacuum wave functional
(\ref{vacuumevolution2}).

In our analysis the causal structure of quantum field theory in
the Functional Schr\"odinger picture is manifest. We have shown
that the causal propagator, given by the vacuum expectation value
of the commutator, forms the essential function in terms of which
the functional kernel is expressed. Therefore, our Functional
Schr\"odinger picture analysis reproduces the standard (Heisenberg
picture) causality structure of FLRW spacetimes.


\begin{thebibliography}{99}
\bibitem{Guth:1985ya}
  A.~H.~Guth and S.~Y.~Pi,
  The Quantum Mechanics of the Scalar Field in the New Inflationary
  Universe,
  Phys.\ Rev.\  D {\bf 32} (1985) 1899.
\bibitem{Prokopec:2006fc}
  T.~Prokopec and G.~I.~Rigopoulos,
  Decoherence from Isocurvature Perturbations in Inflation,
  JCAP {\bf 0711} (2007) 029
  [arXiv:astro-ph/0612067].
\bibitem{Koksma:2007zz}
  J.~F.~Koksma,
  Decoherence of Cosmological Perturbations,
  M.Sc. Thesis (2007),\\
  {\tt http://www1.phys.uu.nl/wwwitf/Teaching/Thesis.htm}.
\bibitem{Weinberg:2005vy}
  S.~Weinberg,
  Quantum Contributions to Cosmological Correlations,
  Phys.\ Rev.\  D {\bf 72} (2005) 043514
  [arXiv:hep-th/0506236].
\bibitem{Prokopec:2003tm}
  T.~Prokopec and E.~Puchwein,
  Photon Mass Generation during Inflation: de Sitter Invariant
  Case,
  JCAP {\bf 0404}, 007 (2004)
  [arXiv:astro-ph/0312274].
\bibitem{Peskin:1995ev}
  M.~E.~Peskin and D.~V.~Schroeder,
  An Introduction to Quantum Field Theory,
  Westview Press (1995).
\bibitem{Jackiw:1988sf}
  R.~Jackiw,
  Analysis on Infinite Dimensional Manifolds: Schroedinger Representation
  for Quantized Fields,
  presented at Seminar on Higher Mathematics, Montreal, Canada, Jun
  1988.
\bibitem{Jackiw:1987aq}
  R.~Jackiw,
  Functional Representations for Quantized Fields,
  published in ``1st Asia Pacific Workshop on High Energy Physics,''
  Sinagpore, Jun 21, 1987 and 6th Symp. on Theoretical Physics,
  Seoul, Korea, Jul 1987 and Mathematical Quantum
  Field Theory, Montreal, Canada, Sep 1987.
\bibitem{Guven:1987bx}
  J.~Guven, B.~Lieberman and C.~T.~Hill,
  Schr\"odinger Picture Field Theory in Robertson-Walker Flat
  Spacetimes,
  Phys.\ Rev.\  D {\bf 39}, 438 (1989).
\bibitem{ProkopecRigopoulos:2007}
  T.~Prokopec and G.~Rigopoulos,
  in progress.
\bibitem{Chernikov:1968zm}
  N.~A.~Chernikov and E.~A.~Tagirov,
  Quantum Theory of Scalar Fields in de Sitter Space-time,
  Annales Poincare Phys.\ Theor.\  A {\bf 9} (1968) 109.
\bibitem{Garbrecht:2006jm}
  B.~Garbrecht and T.~Prokopec,
  Fermion Mass Generation in de Sitter Space,
  Phys.\ Rev.\  D {\bf 73}, 064036 (2006)
  [arXiv:gr-qc/0602011].
\bibitem{Long:1996wf}
  D.~V.~Long and G.~M.~Shore,
  The Schr\"odinger Wave Functional and Vacuum States in Curved
  Spacetime,
  Nucl.\ Phys.\  B {\bf 530}, 247 (1998)
  [arXiv:hep-th/9605004].
\bibitem{Berges:2004yj}
  J.~Berges,
  Introduction to Nonequilibrium Quantum Field Theory,
  AIP Conf.\ Proc.\  {\bf 739} (2005) 3
  [arXiv:hep-ph/0409233].
\bibitem{Nomoto:1992cp}
  K.~Nomoto and R.~Fukuda,
  Causality in the Schr\"odinger Picture: Diagrammatical Approach,
  Phys.\ Rev.\  D {\bf 46}, 1680 (1992).
\bibitem{Hegerfeldt:1974qu}
  G.~C.~Hegerfeldt,
  Remark on Causality and Particle Localization,
  Phys.\ Rev.\  D {\bf 10} (1974) 3320.
\bibitem{Hegerfeldt:1985fy}
  G.~C.~Hegerfeldt,
  Violation of Causality in Relativistic Quantum Theory?,
  Phys.\ Rev.\ Lett.\  {\bf 54} (1985) 2395.
\bibitem{Hegerfeldt:1998ar}
  G.~C.~Hegerfeldt,
  Instantaneous Spreading and Einstein Causality in Quantum Theory,
  Annalen Phys.\  {\bf 7} (1998) 716
  [arXiv:quant-ph/9809030], and references therein.
\bibitem{Halvorson:2001hb}
  H.~Halvorson and R.~Clifton,
  No Place for Particles in Relativistic Quantum Theories?,
  Phil.\ Sci.\  {\bf 69} (2002) 1
  [arXiv:quant-ph/0103041].
\bibitem{Halvorson:2006wj}
  H.~Halvorson and M.~Muger,
  Algebraic Quantum Field Theory,
  [arXiv:math-ph/0602036].
\bibitem{FernandoPerez:1976ib}
  J.~Fernando Perez and I.~F.~Wilde,
  Localization and Causality in Relativistic Quantum Mechanics,
  Phys.\ Rev.\  D {\bf 16} (1977) 315.
\bibitem{Sakurai}
  J.~J.~Sakurai,
  Modern Quantum Mechanics,
  Addison Wesley Longman (1994).
\end{thebibliography}
\end{document}